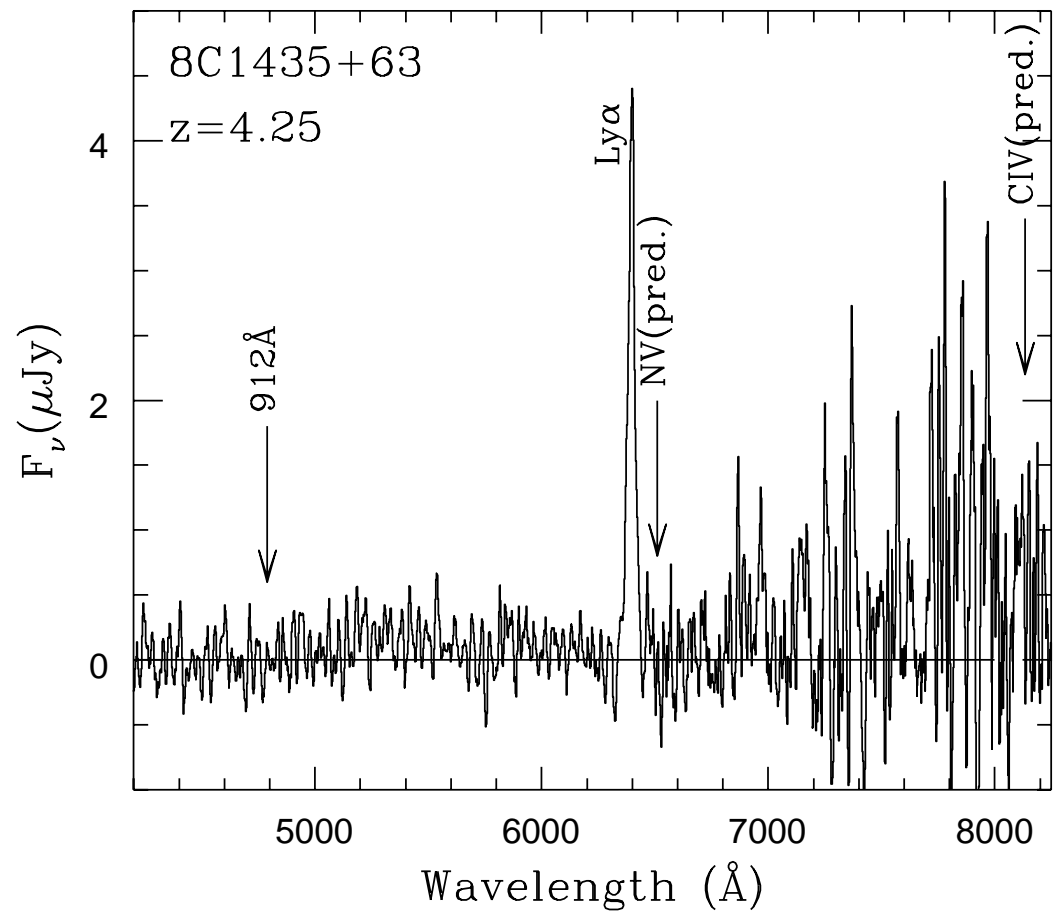

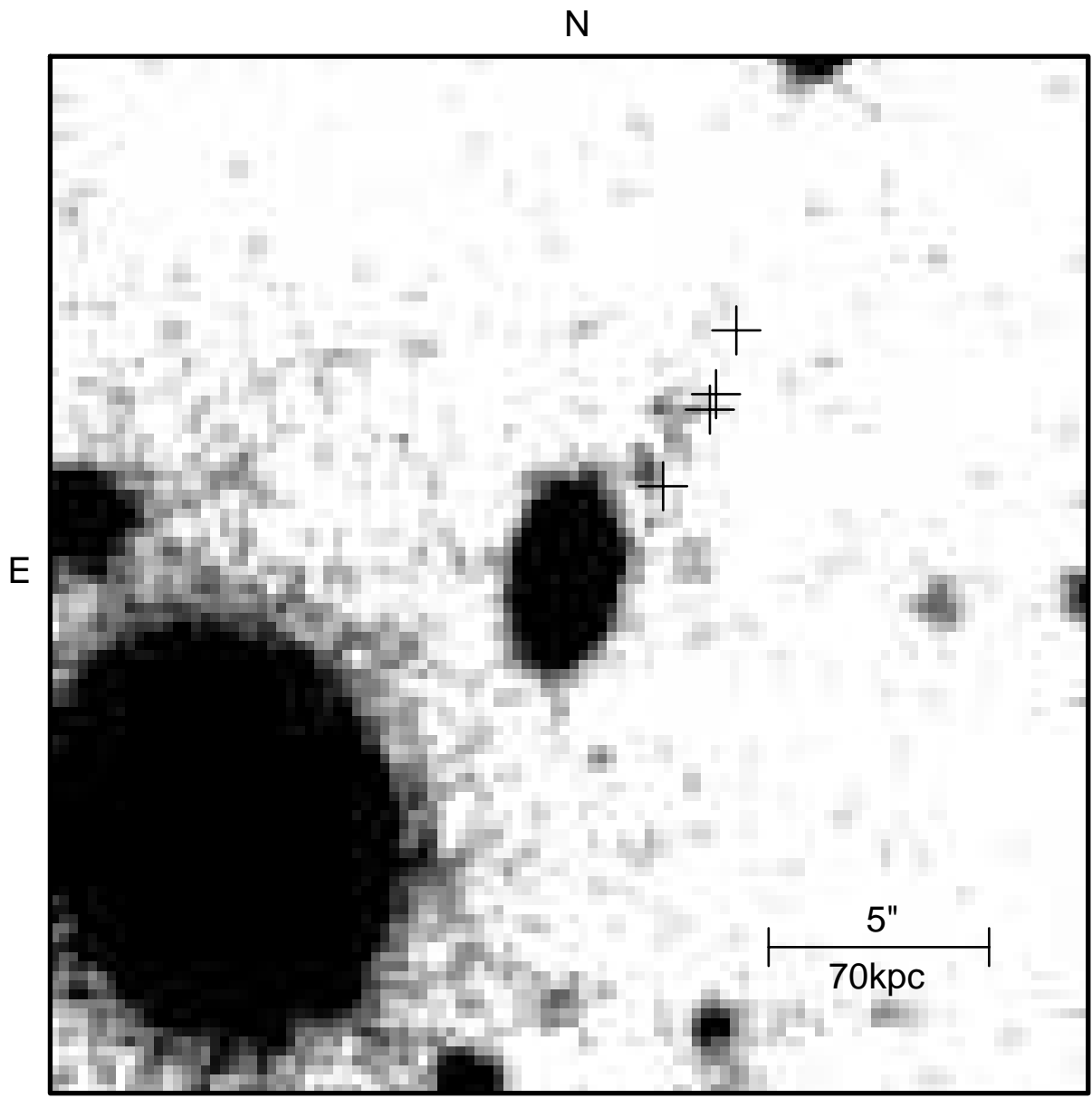

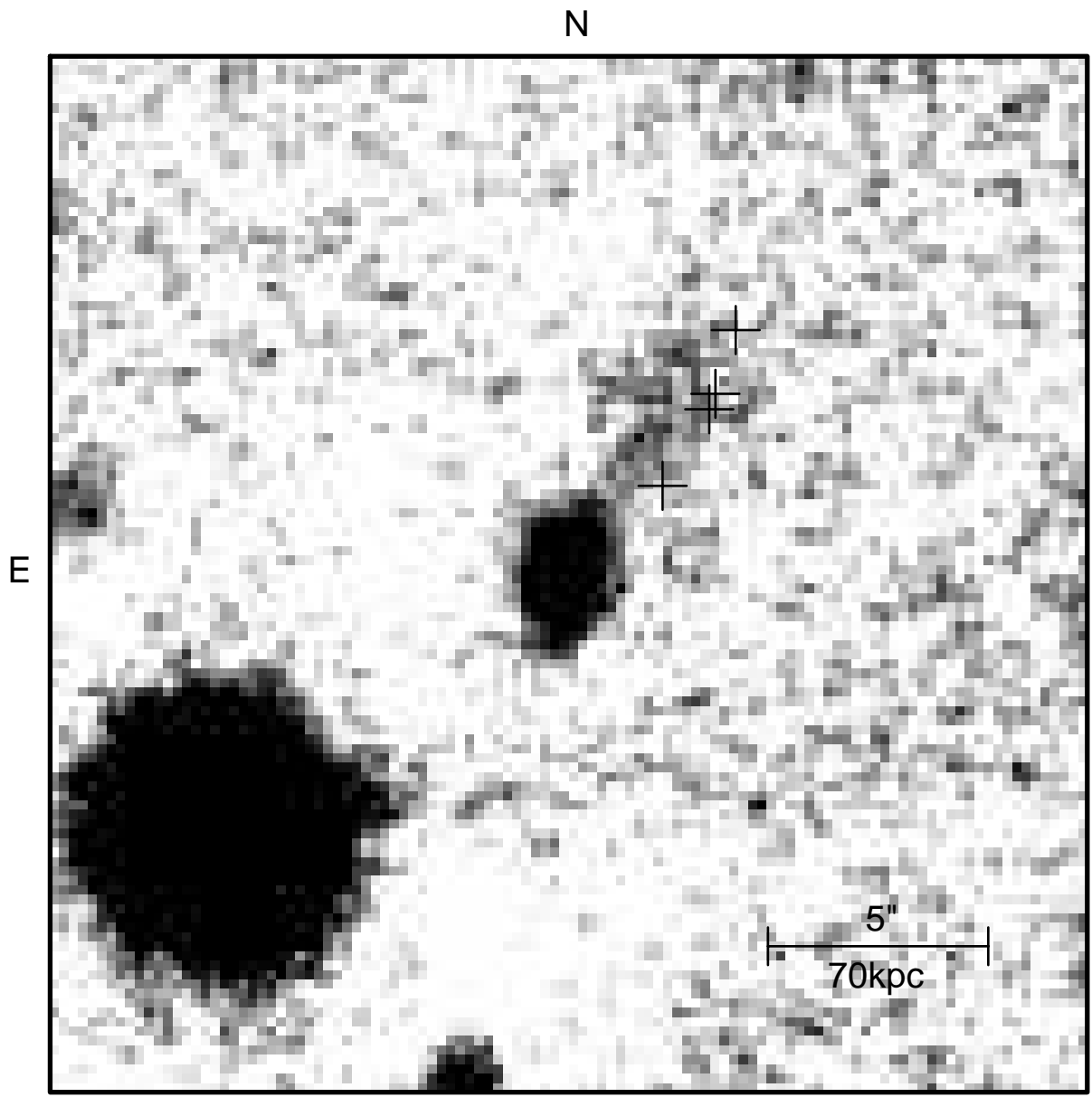

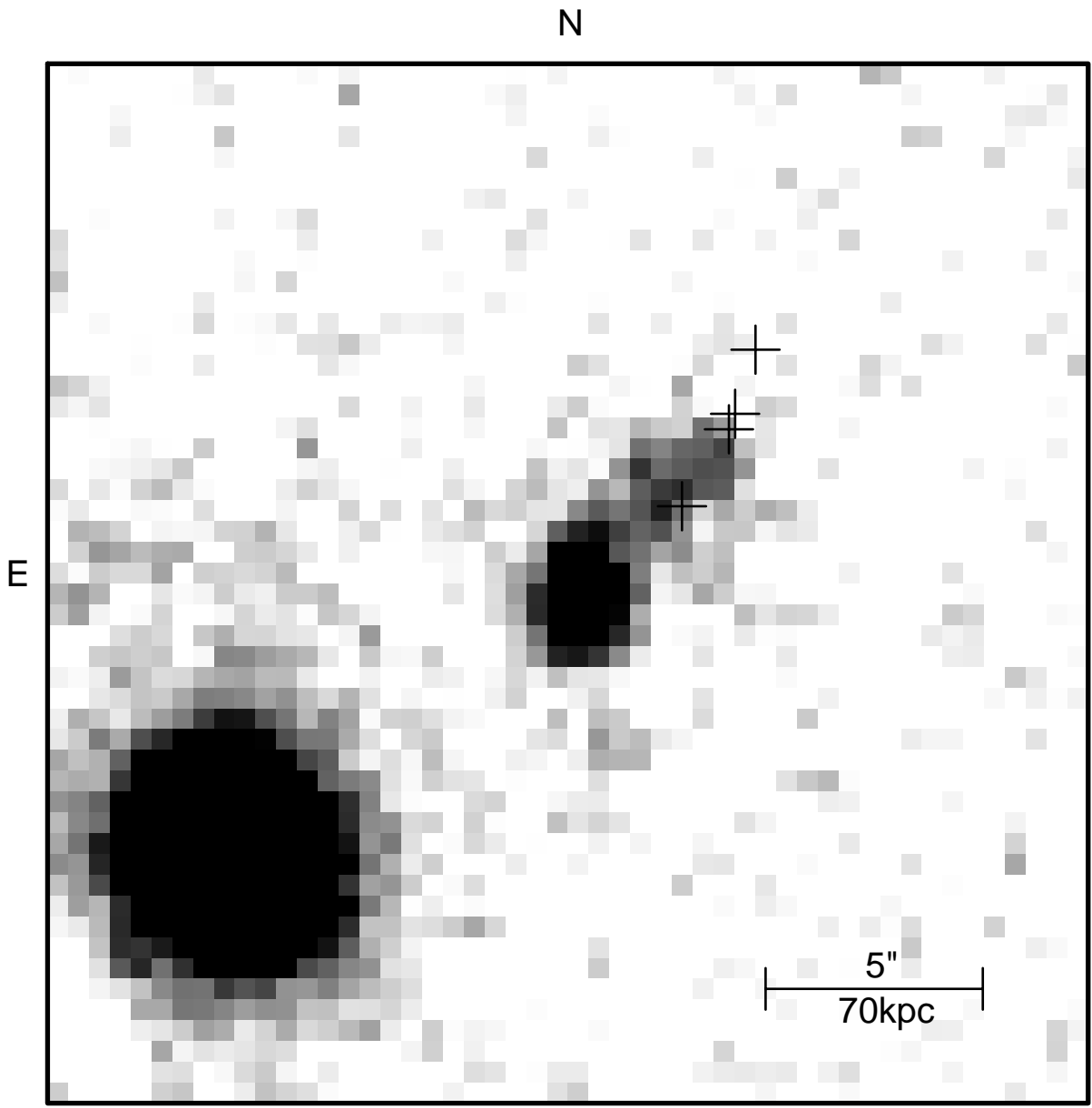

# KECK OBSERVATIONS OF THE MOST DISTANT GALAXY: 8C1435+63 AT z=4.25[1]

HYRON SPINRAD[2], ARJUN DEY & JAMES R. GRAHAM[3]
Astronomy Dept., 601 Campbell Hall, U. C. Berkeley, CA 94720
*Written 1994 September 6*

## ABSTRACT

We report on Keck observations and confirm the redshift of the most distant galaxy known: 8C1435+63 at z=4.25. The spectrum shows a strong Ly$\alpha$ line, a Ly$\alpha$ forest continuum break and a continuum break at $\lambda_{rest} = 912$Å. The Ly$\alpha$ emission is spatially extended and roughly aligned with the radio source. The galaxy shows a double structure in the $I$-band ($\lambda_{rest} \approx 1500$Å) which is aligned with the radio axis; the two $I$-band components spatially coincide with the nuclear and southern radio components. Some fraction of the $I$ band emission could be due to a nonthermal process such as inverse compton scattering. In the $K$-band ($\lambda_{rest} \approx 4200$Å), which may be dominated by starlight, the galaxy has a very low surface brightness, diffuse morphology. The $K$ morphology shows little relationship to the radio source structure, although the major axis of the $K$ emission is elongated roughly in the direction of the radio source axis. The galaxian continuum is very red ($I - K > 4$) and if the $K$ continuum is due to starlight, implies a formation redshift of $z_f > 5$. We speculate that this galaxy may be the progenitor of a present day cD galaxy.

*Subject headings:* cosmology: early universe — galaxies: active — galaxies: redshifts — galaxies: individual (8C1435+63) – galaxies: evolution – radio continuum: galaxies

## 1. INTRODUCTION

The recent discovery of a redshift 4.25 radio galaxy 8C1435+63 by Lacy et al. (1994; hereafter L94) is an important step in the study of the early universe. As the most distant galaxy presently known, 8C1435+63 provides an estimate of the epoch of galaxy formation by using its starlight as a chronometer.

In this paper we present new observations of 8C1435+63. We confirm the redshift and discuss the morphology of the galaxy and its continuum colors. Throughout this paper we use $H_o = 50$ km s$^{-1}$Mpc$^{-1}$ and $q_o = 0$ which results in an angular scale of 14.0 kpc/″ at z=4.25. The lookback time for these parameters is 15.8 Gyr, or 81% of the age of the universe. For reference with $H_o = 75$ km s$^{-1}$Mpc$^{-1}$ and $q_o = 0.5$ the scale is 4.16 kpc/″ and the lookback time is 90% of the age of the universe.

## 2. OBSERVATIONS

We obtained optical imaging and spectroscopic observations with the Low Resolution Imaging Spectrometer (LRIS) of the 10m W. M. Keck telescope on Mauna Kea on U.T. 1994 March 14,15 and July 10. The detector is a dual amplifier read Tek 2048$^2$ CCD with a scale of 0″.214 pix$^{-1}$. There were several problems with the performance of the instrument and detector during the March run (the read noise of both amplifiers was excessively high) and hence more weight is given to our July spectroscopic data. The typical seeing was 0″.8 – 1″.0.

We obtained 5 images of 8C1435+63 (total $t_{exp} = 1000s$) through a long-pass $I$ filter ($\Delta\lambda \approx 2400$Å, $\lambda_o \approx 8700$Å). The images were corrected for overscan bias, flattened using a median sky flat, coadded, and calibrated using observations of the Landolt (1992) standard star field PG1525−071. LRIS has a field of view of $\sim 5'.7 \times 7'.3$ and all the stars in the Landolt field

Table 1: Observed Spectral Lines and Their Fluxes

| Line | $\lambda_{obs}$ Å | Redshift | $F_{obs}^{\dagger}$ | $W_{\lambda}^{obs}$ Å | FWHM$_{obs}$ Å |
|---|---|---|---|---|---|
| Ly$\alpha\lambda$1216 | 6397±1 | 4.262±0.001 | 1.5±0.1 | 3500±500 | 35±2 |
| NV$\lambda$1240 | – | – | <0.15 | – | – |
| CIV$\lambda$1549 | – | – | <0.23 | – | – |

† In units of $10^{-16}$ erg s$^{-1}$ cm$^{-2}$.

were imaged in a single frame; 3 stars were in the linear count regime.

Five 1500$s$ exposure spectra (2 in March, 3 in July) of 8C1435+63 in the wavelength range 3900–8800Å were obtained using a 300 l/mm grating and a 1″ slit (effective resolution FWHM $\approx$ 10Å). The slit was parallel to the major axis of the galaxy (PA=156°); it was also parallel to the average parallactic angle during our observation. Hence, although conditions were non-photometric, our relative spectrophotometry is accurate for $\lambda < 7300$Å. The data were corrected for overscan bias, flat fielded using internal lamps taken after the observations and flux calibrated using observations of the spectrophotometric standards G193-74 (Oke 1990) and HZ44 (Massey et al. 1988). The standard stars were observed both with and without an OG570 filter in order to correct for the second order light contaminating the wavelength region $\lambda > 7500$Å.

We obtained Ly$\alpha$ images of 8C1435+63 at the Lick Observatory 3m Shane and KPNO 4m Mayall telescopes. The Lick data were obtained on 1994 April 5 using the Kast imaging spectrometer at the F/17.5 cassegrain focus through a narrow-band filter ($\lambda_o = 6393$Å, $\Delta\lambda_{FWHM} = 75$Å). Three 1200s integrations were obtained in photometric conditions and were calibrated using observations of Feige 34 (Massey et al. 1988). On 1994 May 24 we obtained service observations of 8C1435+63 using the KPNO 4m PFCCD camera. A single 3000s observation was obtained in 1″.5 seeing through the KP808 filter ($\lambda_o = 6410$Å, $\Delta\lambda_{FWHM} = 95$Å). We calibrated the data using

---

Table 2: Aperture Photometry for 8C1435+63

| Component | Aperture Dia. | $F_{Ly\alpha}$ † | $I$ | $K$ | $I - K$ |
|---|---|---|---|---|---|
| Nucleus | 3″ | 1.7±0.2 | 24.0±0.1 | 20.25±0.07 | 3.8 |
|  | 6″ | 3.6±0.3 | 23.6±0.1 | 19.64±0.08 | 4.0 |
| S. radio hotspot | 3″ | 2.1±0.2 | 24.5±0.2 | 20.64±0.11 | 3.9 |
| N. radio hotspot | 3″ | <0.18 | >25.4 | 21.26±0.19 | >4.1 |

† $F_{Ly\alpha}$ in units of $10^{-16}\,\mathrm{erg\,s^{-1}\,cm^{-2}}$.

the Lick 3m observations. We display only the KPNO Ly$\alpha$ image here as it is superior to the Lick data.

8C1435+63 was observed on 1995 May 4 using the 10m W. M. Keck telescope with the Near Infrared Camera (Matthews & Soifer 1994). The camera is equipped with a SBRC 256×256 InSb array. The pixel size is 0″.15. We observed at K (2.0 – 2.4 μm). An integration time of 5 – 8s per frame was used, and a summed integration time of 120s was obtained per telescope position. All frames were sky subtracted and then flat fielded using dark subtracted sky frames. The positions of two to three objects in each frame were used for spatial registration. The exposure time for the central 40″ of the resultant mosaic is 3480s. Photometry was performed relative to UKIRT faint standards (Casali & Hawarden 1992).

### 3. RESULTS

Figure 1 shows our LRIS spectrum of 8C1435+63. The centroid of the Ly$\alpha$ line has a redshift z=4.261±0.001; the peak of the line is at z=4.264. There is also a break in the continuum at $\lambda_{obs} \approx 4800 \pm 50$ Å which corresponds to the Lyman limit (912Å) at z=4.26±0.05. The flux in the Ly$\alpha$ line determined from a 2″.1 spectral extraction is $1.5\pm0.2 \times 10^{-16}$ erg s$^{-1}$ cm$^{-2}$. The continuum discontinuity across the Ly$\alpha$ line, defined as the ratio of the flux density (in $F_\nu$) between $\lambda\lambda 1250-1350$ to the average flux density between $\lambda\lambda 1100-1200$ is measured to be 2.5±1. The Lyman limit discontinuity ($<f_\nu(970\text{Å})>/<f_\nu(850\text{Å})>$) is greater than 10. We do not detect CIV or NV in our spectrum; the limits on the line ratios relative to Ly$\alpha$ are CIV/Ly$\alpha$ <0.15 and NV/Ly$\alpha$ <0.1. The continuum magnitude at $\lambda\lambda 6600$–7100 ($\lambda_{rest} \approx 1260 - 1350$ Å) is $m_{\lambda 6850} \approx 25$ AB mag.

Figure 2 shows the $I$-band ($\lambda_{rest} \sim 1700$Å) image of 8C1435+63. The composite frame reaches a 1$\sigma$ limiting magnitude of 26.4 in a 3″ diameter aperture. The $I$-band morphology of the galaxy is double, showing a compact component (FWHM $\lesssim$ 0.8″) to the SE and a more diffuse component (FWHM $\sim$ 1″.3) to the NW. Astrometry of the radio source (L94) implies that the compact SE component is coincident with the SE lobe of the radio source whereas the two core components are associated with the more diffuse NW component (Fig. 2). There is no $I$ continuum emission associated with the northern hotspot of the radio source. We subtracted an elliptical model fit to the light distribution of the foreground galaxy ("a") from the $I$ image in order to determine the brightness of 8C1435. The magnitudes of the various components are listed in Table 2. The galaxy is very faint: its $I$ magnitude in a 6″ diameter aperture is 23.6 (or ≈0.9μJy).

The $K$ mosaic ($\lambda_{rest} \sim 4200$Å) achieves a 1$\sigma$ limiting magnitude of 23.6 in a 3″ diameter beam. The image size is seeing limited with FWHM≈1″.1. The $K$ image (Fig. 3) shows that 8C1435+63 consists of an extended region of diffuse emission with a size of $\simeq$ 2″.5 or 35 kpc. Viewing the $K$ image of 8C1435+63 at high contrast suggests that is elongated in a SE – NW direction. However, there is insufficient signal-to-noise to determine if any signicant internal structure has been resolved. In order to measure the $K$ flux of 8C1435+63, we scaled and subtracted the $I$ band model fit to the foreground galaxy. In the same 6″ aperture, $K$=19.6 (or ≈9μJy).

Figure 4 shows our KPNO Ly$\alpha$ narrow band image of 8C1435+63. An off-band continuum image has not been subtracted, but the observed equivalent width of the line is ~3500±500Å and therefore line emission dominates the measured flux through the filter. The Ly$\alpha$ emission is aligned with the radio source and roughly spatially coincident with the $I$-band emission. There is no line emission associated with the northern radio hotspot. The flux in the Ly$\alpha$ line in a 6″ diameter aperture is $3.6\times10^{-16}$erg s$^{-1}$ cm$^{-2}$.

### 4. DISCUSSION

#### 4.1. Spectrum

We confirm that 8C1435+63 is at z=4.25. This is based on the detection of the Ly$\alpha$ line, the continuum break across the Ly$\alpha$ line, the Lyman limit discontinuity at 912Å and the faintness of the galaxy in $I$ and $K$. If, instead, the strong line is [OII]$\lambda$3727 at z=0.71 we would certainly have detected [OIII] and/or H$\beta$, and the $K$ magnitude would have been ≈17±1. Although NV$\lambda$1240 is not detected in our spectrum, the limit (NV/Ly$\alpha$ <0.1) is not inconsistent with that seen in the average radio galaxy spectrum (McCarthy 1993). The CIV$\lambda$1549 emission line reported by L94 is not detected in our spectrum and this allows us to place a limit of CIV/Ly$\alpha$ < 0.15; radio galaxies typically exhibit CIV/Ly$\alpha$ ~0.1.

**Figure 1:** Keck LRIS spectrum of 8C1435+63. The Ly$\alpha$ line and the predicted wavelengths of the NV$\lambda$1240 and CIV$\lambda$1549 emission lines are labelled. Note the break in the continuum at the Ly$\alpha$ line and the lack of flux below ≈4800Å. The continuum magnitude at $\lambda$ ≈7000Å is ≈25 AB mag.

The continuum discontinuity at the Lyα line of ≈2.5±1 is fairly typical of the highest redshift objects (it is ~2 for 4C41.17, the second highest redshift radio galaxy at z=3.8; Chambers, Miley & van Breugel 1990). Just shortward of the Lyα line the continuum is depressed, and then has a broad maximum around $\lambda_{rest} \sim 1050$Å. This is similar to the continuum shape observed in 4C41.17 and the quasar PC 1247+3406 (z=4.9; Schneider, Schmidt & Gunn 1991b), and may be suggestive of Lyα absorption in the vicinity of the object. The Lyα discontinuity can also be described by the broad-band "flux-deficit" parameter usually measured for quasars $D_A = <1 - \frac{f_\nu(\lambda 1050-1170)_{obs}}{f_\nu(\lambda 1050-1170)_{pred.}}>$ (Oke & Korycanski 1982). We measure $D_A \approx 0.45 \pm 0.1$ which is somewhat smaller, but comparable to that seen in z>4 quasars (Schneider, Schmidt & Gunn 1991a,b).

The shape of the Lyα line is consistent with L94's observation of multiple velocity components. In particular, the southern component may be blueshifted relative to the nuclear component by ~ 700 kms$^{-1}$. The restframe equivalent width of the emission line is ~700Å. The large velocity width of the line (~650 kms$^{-1}$) and its spatial coincidence with the radio components argue that the line emission is probably powered by the AGN.

The Lyα line emission is roughly aligned with the radio continuum emission and is associated primarily with the nuclear and southern radio components (Fig. 4). There is little or no line emission associated with the northern radio component. L94 report the detection of radio polarization in the southern lobe; this is unusual since most z~1 radio galaxies generally have line emitting gas associated with the depolarized lobe (Pedelty et al. 1989). However, the northern lobe is fainter than the southern lobe, and its fractional polarization may be larger than that of the southern lobe but still below the sensitivity limit.

### 4.2. Morphology

The broad band images in $I$ and $K$ are both free of strong line emission and correspond to bands centered on 1500Å and 4200Å in the restframe of the galaxy. CIVλ1549 is in the $I$ band, but the line is weak and therefore has a negligible contribution (<4%) to the flux in the band. The only strong lines in $K$ are [NeIII], Hγ and higher order Balmer lines. These lines may contribute slightly in regions where there is Lyα emission, but it is unlikely that they contribute over the full extent of the galaxy.

The restframe UV continuum and Lyα line emission of the galaxy are aligned with the radio axis in PA=156° and in this respect the galaxy is similar to most z>1 radio galaxies (McCarthy et al. 1987, Chambers et al. 1987). Since the UV emission coincides roughly with components of the radio source, it is possible that the some fraction of this continuum emission is of nonstellar origin.

The $I$ and $K$ continuum fluxes cannot be due to synchrotron emission unless $\alpha^{radio}_{optical} \approx 0.75$ which is much flatter than the observed high radio frequency spectral index $\alpha^{8.4\text{GHz}}_{15.2\text{GHz}} = 2.1$ (L94). In analogy to lower redshift radio galaxies, some of the $I$ light may be light from a hidden AGN scattered by dust or electrons. Polarization data is necessary to address this, but we note here that a large fraction of the $I$ flux may be explained by inverse Compton (IC) scattering of cosmic microwave background photons by γ ~ 25 − 100 relativistic electrons in the synchrotron emitting plasma (Daly 1992). The radio spectrum of 8C1435+63 flattens at low frequencies and we can obtain an estimate for the IC scattered flux ($f_{IC}$) by considering a range in spectral index 0.27 > α > 0 where the upper bound is $\alpha^{38\text{MHz}}_{151\text{MHz}}$. This predicts IC scattered $I$ and $K$ flux densities of $2.5 B_\perp^{-1.27} \mu\text{Jy} > f^I_{IC} > 0.09 B_\perp^{-1} \mu\text{Jy}$ and $3.2 B_\perp^{-1.27} \mu\text{Jy} > f^K_{IC} > 0.09 B_\perp^{-1} \mu\text{Jy}$, where $B_\perp$ is the magnetic field (in μG) in the radio lobe (Daly 1992). Hence it is possible that some fraction of the $I$-band continuum may be IC scattered light. 8C1435+63 is unique in this respect: the large energy density of the cosmic microwave background at z>4 and the radio galaxy's large low frequency flux density make it one of the few cases where IC scattering is likely to be a component of the restframe UV continuum.

In contrast to the restframe UV morphology, the morphology at $\lambda_{rest} \sim 4200$Å is much more diffuse and extended: the FWHM of the $K$ image is ~2″.5 compared to ~1″.3 for the $I$ image. The galaxy has extremely low $K$ surface brightness ($\mu_K \approx 22.5$ mag/□″) and has a much more uniform morphology than in the $I$ band. It is roughly extended along the radio axis, but unlike the $I$ band, the $K$ band morphology is not obviously related to the radio source. Since (1) the radio spectral index is undoubtedly very flat (or possibly inverted) at the low frequencies (~100 Hz) necessary to produce IC scattering at $K$ wavelengths and (2) the $K$ morphology bears little resemblance to the radio morphology, it is very unlikely that any significant component of the $K$ light is due to IC scattering.

The $K$-band morphology of this object is different from that of other high redshift radio galaxies. The second most distant galaxy known, 4C41.17 (z = 3.8, Chambers, Miley & van Breugel 1990) shows a close spatial correspondence between the optical, infrared and radio emission components (Graham et al. 1994). The z=3.22 radio galaxy 6C1232+39 (Eales et al. 1993) has a compact, nucleated infrared morphology. The four z>2 objects imaged by McCarthy, Persson & West (1992) have infrared morphologies that are either nucleated or similar to their optical morphologies. The case of B2 0902+34 (z=3.4, Lilly 1988) is more complicated; the $K$ band light is largely due to [OIII] line emission, but it also may have a large extended low surface brightness continuum 'halo' (Eisenhardt & Dickinson 1992, Eales & Rawlings 1993) similar to that seen in 8C1435+63.

The difference in $I$ and $K$ morphology suggests that different continuum emission mechanisms contribute in the two bands. The $I$ band (restframe UV) emission could have some relation to the emission processes powered by the active nucleus, but the $K$ band light is probably starlight. An important test of this hypothesis will be a measurement of the $H-K$ color which effectively measures the 4000Å break.

The $K$ magnitude implies that the galaxy is slightly brighter by ≈0.5 mag. than the extrapolated $K$–z relationship for radio galaxies (e.g. Eales & Rawlings 1993) but is well within the scatter in the relation. Of course, the relationship is not well determined for very high redshift galaxies and the dispersion is larger at z>2 than at lower redshift (Eales et al. 1993).

### 4.3. Color

Although the $I$ and $K$ emission may have different origins, if the $K$ light is starlight, then the observed $I-K$ color is a lower limit to the true color of the stellar population. The color of the source in a 3″ diameter aperture centered on the nuclear $I$ component is $(I-K) \approx 4$. Since the galaxy is redder at larger radii, this is a lower limit to the integrated $(I-K)$ color. If we assume that the continuum light is stellar in origin, then we may obtain an estimate for the formation redshift ($z_f$) based on models for the evolution of the population. For the passive evolution of a solar metallicity instantaneous burst population formed with a standard IMF, $(I - K) > 4$ implies a formation redshift for the galaxy of $z_f > 5$. If we consider the evolution of a population formed with an IMF with an arbitrary lower mass cutoff of $m_L = 2.5 M_\odot$, then $z_f$ may be as small as 5; a higher $m_L$ would imply a lower estimate for $z_f$ (Charlot et al. 1993).

For $H_o = 50, q_o = 0$, the galaxy must form and obtain an $I - K > 4$ within $\Delta t \approx 3.7$Gyr from the big bang; for $H_o = 75, q_o = 0.5$, $\Delta t < 1$Gyr. The age of the galaxy is $\gtrsim 0.5$Gyr if the color is interpreted in terms of a normal IMF, solar metallicity burst model. Hence if we are observing unreddened starlight it is difficult to reconcile the color of this population with a large value for $H_o$. Note that the assumption of solar metallicity provides an underestimate to the age of the population.

L94 suggest that the galaxy continuum may be reddened by dust based on the fact that the Ly$\alpha$ line is somewhat underluminous compared to other radio galaxies of similar radio power. However, there is substantial scatter in the Ly$\alpha$ luminosity – radio power relationship and 8C1435+63 lies roughly within the scatter (e.g. McCarthy 1993). In addition, if the intrinsic continuum has a flat spectrum ($f_\nu \sim \nu^0, I-K=1.44$), then the observed color implies $E(B-V) > 0.6$ for a Galactic type extinction curve. This is considerably larger than that determined from observations of the Ly$\alpha$/H$\beta$ ratio in other high redshift radio galaxies (McCarthy et al. 1992, Eales & Rawlings 1993) and is more than sufficient to completely suppress the Ly$\alpha$ emission. In any event, a young, dusty interpretation for the galaxian light places a more stringent constraint on the timescale for the formation of the first generation of stars.

Since the $K$ band corresponds to $\lambda_{rest} \approx 4200$Å and the age of the population is of the order of $\sim 0.5$Gyr then the light in this band is mostly dominated by main sequence stars (Charlot and Bruzual 1991). The average $K$ surface brightness corresponds to a $B$ surface luminosity density of $\sim 90 L_\odot$ pc$^{-2}$, or the equivalent of $\sim 2$ A stars per square parsec. This is only a few times larger than the typical surface density of a present day gE/cD galaxies ($\sim 14 L_\odot$ pc$^{-2}$ averaged in a similar metric aperture).

This comparison prompts us to conclude the section on a speculative note. Under the assumption that the $\lambda_{rest} \approx 4200$Å light is primarily starlight, we may guess at the final (i.e. present day) evolutionary product of this galaxy. The passive evolution of a stellar system that is created in an instantaneous burst will fade in the restframe $B$-band by $\approx 2.2$ mag as it ages from $10^9$ to $10^{10}$ Gyr (Charlot & Bruzual 1991). In a 3″ diameter aperture, the average surface brightness of 8C1435+63 is $\mu_K \approx 0.7 \mu$Jy/□″. Assuming that the galaxy fades by 2.2 mag due to evolution and that the monochromatic surface brightness dims as $\mu_\nu \sim (1+z)^{-3}$ due to the expansion of the universe, the present day counterpart would have a surface brightness of $\approx 14 \mu$Jy/□″ or $\mu_B \approx 21.3$ mag/□″ in an aperture of metric diameter 42kpc ($H_o = 50, q_o = 0$). This is roughly consistent with photometry of nearby gE and cD galaxies (Oemler 1976, Sandage & Perelmuter 1990). Note that at the redshift of 8C1435+63 the cosmological dimming term is comparable in magnitude to the evolutionary term. Observations of galaxies at these large redshifts, now possible with the new generation of large telescopes, will eventually allow cosmologically interesting tests (such as the Tolman test; Sandage & Perelmuter 1990,1991). The relative importance of population evolution on these tests will be somewhat reduced because of the increasing significance of the cosmological dimming term at larger redshifts.

We thank M. Lacy and S. Rawlings for generously sharing their results prior to publication and to S. Charlot for stellar population synthesis predictions. We thank T. Bida, W. Wack, J. Aycock, W. Harrison & B. Schaeffer for invaluable help during our Keck runs. We are very grateful to D. DeYoung and R. Green for granting us discretionary observing time at the KPNO 4m on short notice and to A. Sarajedini for obtaining the data for us. We are grateful to J. Najita for a careful reading of the manuscript. H.S. gratefully acknowledges NSF grant # AST-9225133. The W. M. Keck Observatory is a scientific partnership between the University of California and the California Institute of Technology, made possible by the generous gift of the W. M. Keck Foundation.

REFERENCES

Bruzual A., G. & Charlot, S. 1993, ApJ, 405, 538
Casali, M. M. & Hawarden, T. G. 1992, JCMT-UKIRT Newsletter, 4, 33
Chambers, K.C., Miley, G.K. & van Breugel, W.J.M. 1987, Nature, 329, 604
Chambers, K.C., Miley, G.K. & van Breugel, W.J.M. 1990, ApJ, 363, 21
Charlot, S. & Bruzual A., G. 1991, ApJ, 367, 126
Charlot, S., Ferrari, F., Matthews, G.J. & Silk, J. 1993, ApJ, 419, L57
Daly, R.A. 1992, ApJ, 386, L9
Eales, S.A. & Rawlings, S. 1993, ApJ, 411, 67
Eales, S.A. et al. 1993, ApJ, 409, 578
Eisenhardt, P.R.M., & Dickinson, M. 1992, ApJ, 399, L47
Graham, J.R. et al. 1994, ApJ, 420, L5
Lacy, M. et al. 1994, MNRAS, in press. (L94).
Landolt, A.U. 1992, AJ, 104, 340
Lilly, S.J. 1988, ApJ, 333, 161
Massey, P., Strobel, K., Barnes, J.V. & Anderson, E. 1988, ApJ, 328, 315
Matthews, K. & Soifer, B.T. 1994, in Infrared Arrays in Astronomy: The Next Generation, Ed. I.S. McClean, Kluwer:Dordrecht.
McCarthy, P.J. 1993, ARA&A, 31, 639
McCarthy, P.J., Elston, R. & Eisenhardt, P. 1992, ApJ, 387, L29
McCarthy, P.J., Persson, S.E. & West, S.C. 1992, ApJ, 368, 52
McCarthy, P.J., van Breugel, W.J.M., Spinrad, H. & Djorgovski, S. 1987, ApJ, 321, L29
Oemler, A. 1976, ApJ, 209, 693
Oke, J.B. 1990, AJ, 99, 1621
Oke, J.B. & Korycanski, D.G. 1982, ApJ, 255, 11
Pedelty, J.A., Rudnick, L., McCarthy, P.J. & Spinrad, H. 1989, AJ, 97, 647.
Sandage, A. & Perelmuter, J.-P. 1990, ApJ, 350, 481
Sandage, A. & Perelmuter, J.-P. 1991, ApJ, 370, 455
Schneider, D.P., Schmidt, M. & Gunn, J.E. 1991a, AJ, 101, 2004
Schneider, D.P., Schmidt, M. & Gunn, J.E. 1991b, AJ, 102, 837

**Figure 2:** Keck LRIS $I$-band image of 8C1435+63. The crosses mark our best estimate for the positions of the four radio components shown in Lacy *et al.* 1994. Note the absence of any emission associated with the northern radio component. The centroid of the nuclear component is at $\alpha_{1950} = 14^h35^m27\overset{s}{.}5, \delta_{1950} = +63°32'12\overset{''}{.}6$. The bright galaxies to the SE are members of a foreground group at $z \approx 0.23$. The $5''$ scale bar in the lower right corresponds to 70kpc for $H_o = 50\,\mathrm{km\,s^{-1}\,Mpc^{-1}}$, $q_o = 0$.

**Figure 3:** Keck NIRC $K$-band image of 8C1435+63. Note that the image has a diffuse, extended structure. In contrast to the $I$ image, there is continuum emission north of the nuclear radio components. The $5''$ scale bar in the lower right corresponds to 70kpc for $H_o = 50\,\mathrm{km\,s^{-1}\,Mpc^{-1}}$, $q_o = 0$.

**Figure 4:** KPNO 4m narrow band Ly$\alpha$ image of 8C1435+63. Note that the line emission is primarily associated with the southern and nuclear components of the radio source. There is no line emission associated with the northern radio component.